\documentclass[doublecol,linenumbers]{epl2}
\usepackage{graphicx}
\usepackage[breaklinks=true,colorlinks=true,linkcolor=blue,urlcolor=blue,citecolor=blue]{hyperref}
\RequirePackage[T1]{fontenc}
\usepackage[british]{babel}
\RequirePackage{mathptmx}      
\RequirePackage{flushend}
\usepackage{cite}
\usepackage{amsmath,amsfonts,amssymb}
 \expandafter\let\csname equation*\endcsname\relax
  \expandafter\let\csname endequation*\endcsname\relax
\usepackage[mathcal]{eucal}
\usepackage{mathrsfs}
\usepackage{latexsym}
\usepackage{bm}
\usepackage[all]{xy}
\usepackage{dsfont}
\usepackage{upgreek}

\title{Fine-tuning the Color-Glass Condensate with the nuclear configurational entropy}
\shorttitle{Fine-tuning the Color-Glass Condensate} 

\author{G. Karapetyan \thanks{E-mail:\email{gayane.karapetyan@ufabc.edu.br}}}
\shortauthor{G. Karapetyan}

\institute{Centro de Ci\^encias Naturais e Humanas, Universidade Federal do ABC - UFABC\\ 09210-580, Santo Andr\'e, Brazil.}                   


\title{Fine-tuning the Color-Glass Condensate with the nuclear configurational entropy}

\abstract{
The dipole-nucleus forward scattering amplitude rules the onset of the gluon anomalous dimension, in the Color-Glass Condensate regime. In this model, the onset of quantum regime is here \emph{derived} as a critical stable point in the nuclear configurational entropy, matching the fitted experimental data in the literature  with accuracy of $\sim1\%$. It corroborates with the informational entropy paradigm in high energy nuclear physics. 
}

\begin{document}
\maketitle

\section{Introduction}
 In the specific case of high energy nuclear physics, the informational entropy setup has been brought out into a multiplicity of scopes in hadronic processes, in the lattice QCD setup, and in the AdS/QCD correspondence as well \cite{Bernardini:2016hvx, Bernardini:2016qit,Casadio:2016aum,ads}. In these contexts, the stability of mesons \cite{Bernardini:2016hvx} and the dominance of glueball states \cite{Bernardini:2016qit} were derived, in the context of the configurational entropy.  
 The informational entropy has its roots in the Shannon work of communication theory, measuring the shape complexity of spatially-localized configurations, as the expected value of the information contained in each message, in communication theory. 
The concept of the informational entropy was refreshed by means of the relative configurational entropy  
\cite{Gleiser:2012tu,Sowinski:2015cfa,Gleiser:2011di,Gleiser:2015rwa}, that up to now has relied on the energy density that represents the studied systems. Both denominations, configurational and informational entropy, are currently used throughout the literature and hereupon we also utilize both namings. The informational entropy counts on the Fourier transform 
of  square-integrable, 
bounded, position-dependent functions. Such rich concept computes the inherent shape complexity related to configurations of physical systems that are spatially-localized. Here we propose the informational entropy as a quite natural tool for the study  of 
cross sections in nuclear physics. In fact, the cross sections are also physically adequate to define the configurational entropy,  since they are also spatially-localized, square-integrable, position-dependent 
bounded functions. There are close parallels between the mathematical expressions for the thermodynamical entropy and the informational  entropy. As the information entropy has been calculated for energy densities, the choice of the total reaction cross section can be, therefore,  as well as appropriate for this purpose.
In fact, the amount of information needed to define the detailed quantum states of the system, given its macroscopic description, can be thus implemented.

For configurations encoded 
by the the measure of their cross section in nuclear physics, information entropy can be thought of as being a measure of the weights associated with different modes, in the momentum space, that comprehends the nuclear physical system to be studied.  The
higher the informational entropy, the higher the information, necessary to portray the shape, is. 
Cross sections in the Color-Glass Condensate (CGC) are here analysed,  with respect to the critical points of the associated informational entropy.

\section{The configurational entropy and KKT model}

In the last decades, the mechanism of hadron interactions at high energy regime within Quantum Chromodynamics (QCD) has been called the attention of particle physicists. The dependence of the total hadronic cross section with the energy permits to relate such aspect to the reaction mechanism based on QCD, in particular, a suggestion of the appropriate role of the partons in the high energy reactions with hadrons. 
QCD at high energies can be described as a many-body theory of partons which are weakly coupled albeit non-perturbative due to the large number of
partons. 

The so-called Color-Glass Condensate is the effective theory describing high energy scattering in QCD. It provides insight into outstanding conceptual issues in QCD at asymptotically high energies. For instance, in the case of Deep Inelastic Scattering (DIS) of a lepton scattering off a hadronic target, 
it  provides the useful environment to study the particle production at high energies. During the collision one can represent
the hadron as a collection of constituents, partons, which are nearly on-shell excitations carrying some fraction $x$ of the total hadron longitudinal momentum.
This $x$ variable is equal to the empirical Bjorken variable $x_{Bj}$. 
The inclusive cross sections in DIS can be expressed in terms of the
Lorentz invariants, such as $x$ which corresponds, at lowest order in perturbation theory,
to the longitudinal momentum fraction carried by a parton in the hadron, the virtual photon four-momentum squared $q^{2}= -Q^{2} < 0$ exchanged between the
electron and the hadron, and the center of mass energy
squared $s$.
 The CGC has been proposed as a new state of matter, characterized by gluon saturation and by a typical momentum (saturation) scale, $Q_s$, growing with the energy. The Color-Glass Condensate  determines the critical interface that separates the linear and the saturation regimes of QCD dynamics \cite{cgc,kkt,Carvalho,Carvalho:2008zzb}. Ref. \cite{kkt} considered a sea of gluons in gold ions that  seem to other ions as a gluon wall, constituting the Color-Glass Condensate itself, comprising gluons of high density. The higher the energy, the higher the momentum states -- that are occupied by
the gluons -- are, and concomitantly the weaker the coupling among the gluons is \cite{cgc,kkt,Carvalho}. Hence, the gluon density saturates, corresponding
to a multiparticle Bose condensate state. Ref. \cite{Casadio:2016aum} thoroughly 
paves the road for this interpretation in the informational entropy setup.

In the QCD framework, the total hadronic cross section has been shown to  be correctly described by the minijet model, with the inclusion of a window in the $p_{T}$-spectrum associated to the saturation physics \cite{Carvalho}. This model was also successfully used for the dipole-nucleus forward scattering amplitude, which described the onset of the gluon anomalous dimension in the color glass condensate regime \cite{kkt}.
Partons emulate localized-energy systems, being spatially-coherent field configurations  \cite{Gleiser:2012tu}. The configurational entropy is here proposed to identify both the onset of instability of a spatially-bound configuration as well as the approach towards the optimal state corresponding to the minimum in the configurational entropy. This setup has been widely verified in different contexts, with great 
experimental, phenomenological and observational accuracy (see \cite{Bernardini:2016hvx, Bernardini:2016qit,Casadio:2016aum,ads,Gleiser:2012tu,Sowinski:2015cfa,Gleiser:2011di,Gleiser:2015rwa} and references therein).

The starting point for defining the configurational entropy, generally for $p$ spatial dimensions, is the Fourier transform  of the energy density, given by  
\begin{equation}
{\boldsymbol\rho} ({ \bf k})=(2\pi)^{-p/2}\!\! \overbrace{\int_{-\infty}^\infty\ldots\int_{-\infty}^\infty}^{p\; {\rm times}}  \!\rho({\bf r})\,e^{i{\bf k} \cdot{\bf r}} d^p  r\,\,.
\label{1}
\end{equation} The configurational entropy is defined from the modal fraction, that is constructed upon of the Fourier 
transform of the energy density associated with a  physical system. Then we define the nuclear modal fraction reads:
\begin{equation}
f({\bf k})=\frac{\vert\,\boldsymbol\rho({\bf k})\,\vert^2}{{\int_{-\infty}^\infty\ldots\int_{-\infty}^\infty}\vert\,{\boldsymbol\rho({\bf k})\vert\,^2}\,d^p {k}}.\label{2}
\end{equation}  
 The nuclear configurational entropy is, then,  defined as  \cite{Gleiser:2012tu,Sowinski:2015cfa}
\begin{equation}\label{3}
S_c[f] \,= \, - {\overbrace{\int_{-\infty}^\infty\ldots\int_{-\infty}^\infty}^{p\; {\rm times}}}  f({\bf k} ) \log  f ({\bf k} )  d^p k. 
\end{equation} 
 Critical points of the configurational entropy  imply that the system has informational entropy 
that is critical with respect to the maximal entropy $f_{\rm max}({{\bf k}})$, corresponding to 
 more dominant states \cite{Bernardini:2016hvx,Casadio:2016aum}. It has been comprehensively 
scrutinised and applied in a variety of models that range from 
higher spin mesons and glueballs stability \cite{Bernardini:2016hvx, Bernardini:2016qit} -- in the holographic nuclear 
physics panorama -- to Bose-Einstein condensates of long-wavelength gravitons \cite{Casadio:2016aum}.

On the other hand, the investigation of deep inelastic scattering of leptons on protons and/or nuclei shows that during the interaction the gluons with small $x$, a longitudinal momentum fraction carried by a parton in the hadron, can be produced. The data, obtained in different experiments such as DESY collider HERA, can be successfully interpreted by the phenomenological descriptions based on gluon saturation concept \cite{Dosch,bk,trianta}. It provides an idea that at high energy regimes the nuclei can be considered as systems of gluons with small $x$ value that are strongly correlated. For such dense systems, the theory has been called the Color-Glass Condensate  based on QCD and has been developed in \cite{cgc}. 
At Relativistic Heavy Ion Collider (RHIC) one can observe the window 
for the gluon distributions at 
the energy of a center-of-mass $200$ GeV. For instance, in the deuteron-induced reaction of gold, hadrons, which are detected on the beam direction, are produced mainly by quark-gluon interactions. In such a collision, it is possible to investigate the small $x$ components of the gold nuclei wave function. Such fact leads to the dependence on the number of nucleons and connection with the rapidity $y$ of measured particles by $Q^2_s \sim A^{1/3} e^{\lambda y}$. The value of $\lambda \sim 0.2 - 0.3$ is obtained by fitting the HERA data. 
It should be worth to note that at RHIC energies the saturation scale for gold nuclei is expected to be $\sim$ 2 GeV \cite{cgc}.

In order to compute the informational entropy associated to the Color-Glass Condensate, let us start by calculating the spatial Fourier transform of the
total hadronic cross sections defined in minijet satiration model \cite{Carvalho}. 
Instead of the energy density, we here use the cross sections as a more suitable quantity to describe nuclear systems, being also spatially-localized, square integrable, functions. We will analyze the informational entropy of cross sections in the KKT model. Analogously to the definitions (\ref{1}, \ref{2}), 
the nuclear configurational entropy is hereon defined by means of the Fourier transform of the cross section, 
\begin{equation}\label{34}
{\boldsymbol\sigma} ({ \bf k})=\frac{1}{2\pi}\! \int_{-\infty}^\infty\int_{-\infty}^\infty  \!\sigma({\bf r})\,e^{i{\bf k} \cdot{\bf r}} d {\bf r}\,\,.
\end{equation}  It yields the modal fraction:
\begin{equation}\label{modall}
f_{\boldsymbol\sigma}({\bf k})=\frac{\vert\,\boldsymbol\sigma({\bf k})\,\vert^2}{\int_{-\infty}^\infty\int_{-\infty}^\infty\vert\,{\boldsymbol\sigma({\bf k})\vert\,^2}\,d{\bf k}}.
\end{equation}  
 The configurational entropy is, then,  defined as  \cite{Gleiser:2012tu}
\begin{equation}\label{333}
S_c\left[f_{\boldsymbol\sigma}\right] \,= \, - \int_{-\infty}^\infty\int_{-\infty}^\infty f_{\boldsymbol\sigma}({\bf k} ) \log  f_{\boldsymbol\sigma} ({\bf k})  d {\bf k}. 
\end{equation} 
 Critical points of the nuclear configurational entropy  imply that the system has informational entropy 
that is critical with respect to the maximal entropy $f_{\rm max}({{\bf k}})$, corresponding to 
 more dominant states \cite{Bernardini:2016hvx,Casadio:2016aum,ads}.

This model is based upon the so-called saturation window, that consists of a gap lying in between the non-perturbative and the 
perturbative regimes of QCD. 
In the minijet model \cite{Carvalho} the total hadronic cross section can be decomposed as follows: 
\begin{eqnarray}
\sigma_{\rm total} &=&
\displaystyle{\int_0^{\Lambda^2_{\rm QCD}}} d \sigma  +
\int_{\Lambda^2_{\rm QCD}}^{Q_s^2} d \sigma +
\int_{Q_s^2}^{s/4}  d \sigma  \nonumber  \\
\end{eqnarray} where $d\sigma=\frac{d\sigma}{dp_T^2} dp_T^2$ and  $\sigma_{0}$ characterizes the non-perturbative contribution \cite{Dosch}.
The last split component, 
$\sigma_{\rm pQCD}$, was calculated using the perturbative QCD with respect to a low transverse momenta cut-off. The cross section 
$\sigma_{\rm sat}$ stands for  the saturated component. Such component contains data about interactions below the saturation scale ($Q_{s}$) where the parameter $\sqrt{s}$ fixes the center of mass energy.  It is  
determined by the solution of the non-linear evolution equation
associated to Color-Glass Condensate physics \cite{bk} and given by $\frac{Q_s^2 (x)}{Q_0^2}= \,(\frac{x_0}{x})^{\lambda}$, 
where $x$ is the Bjorken variable, with $Q_0^2=0.3$ GeV$^2$ and $x_0=0.3 \times 10^{-4}$ fixed by the 
initial condition \cite{kkt,Carvalho}. The saturation exponent $\lambda$ has been estimated
on the base of different approaches for the QCD dynamics, being $
\approx 0.3$ \cite{trianta}, agreeing the
HERA  data \cite{123}.

In the above mentioned model \cite{Carvalho}, 
the saturated component is given by the following equation $\sigma_{\rm sat}$ \cite{shoshi}:
\begin{eqnarray}\label{saat}
\sigma_{\rm sat}=
{\iint} 
\sigma_{\rm dip} (x,r)  |\Psi_{p}(r)|^ 2 d^2 r
\end{eqnarray}
where $r$ is the dipole radius and   
the proton wave function $\Psi_p$ is chosen as
\begin{eqnarray}\label{satt}
|\Psi _p(r)|^ 2 =
\frac{1}{2 \pi S^ 2_ p} \, e^{-{r^ 2}/{2 S^ 2 _p}}
\end{eqnarray}
with $S_p=0.74$ fm and the dipole-proton cross section \cite{kkt}
\begin{eqnarray}
\!\!\!\!\!\!\!\!\!\sigma_{\rm dip}(x,r) = 
2 \pi R_p^2 \, \Bigg\{1\!-\! \exp\left[-\frac{1}{4} \left(r\bar{Q}_s\right)^{2\gamma(y,r^2)}\right]\Bigg\}.
\end{eqnarray}
where $R_p = 0.9$ fm and $y$ is the total rapidity interval.\footnote{It is worth 
to mention that the wave function in Eq. (\ref{saat}) can be further 
 generalized in the context of cyclic defects \cite{Chinaglia:2016aat,Bernardini:2014yka,Correa:2015vka}.}
The dipole scattering amplitude should be given by the
impact parameter ($b$) dependent solution of a non-linear evolution equation, such
as the Balitsky-Kovchegov equation \cite{bk}.  
In the KKT model \cite{kkt}, the expression 
for the quark dipole-target scattering amplitude regards 
the relation  $\bar{Q}_s = \pm \frac{2}{3} \, Q_s$  \cite{kkt}, and the anomalous dimension 
reads
\begin{eqnarray}\label{c1}
\!\!\!\!\!\gamma(y, r^2)\! = \!\frac{1}{2}\left(1\!+\!\frac{\xi 
(y, r^2)}{7
\zeta(3)\, c+\xi (y,r^2) \!+\!{2 \,\xi^{1/2} (y, r^2)}} \right),
\end{eqnarray}
with $c$ a free parameter that  describes the onset of quantum regime, which was fixed in to be $c=4$ in the literature \cite{kkt,Carvalho}, and 
\begin{eqnarray}
\xi (y, r^2) \, = -4\, \frac{\log  \vert r Q_{s0} \vert}{\lambda(y-y_0)}\,.
\end{eqnarray}
The saturation scale is expressed as:
$Q_s^2(y)  = \Lambda^2 \sqrt{A}^{3} {x}^{-\lambda}$, where
the parameters $\Lambda=0.6$~GeV and 
$\lambda=0.3$ are fixed by DIS data \cite{GBW}. 
In the theory of the Color-Glass Condensate, the parameter $\lambda$ changes with the energy, being a function 
of the variable $y= \log (1/x)$. In Eq. (12) the initial saturation scale is defined by $Q_{s0}^2=Q_s^2(y_0)$ with $y_0 = 0.6$ being  the
lowest value of rapidity at which the low-$x$ quantum evolution
effects are essential. This parametrization can describe the dAu RHIC data
 \cite{Carvalho,kkt}. 

In general, free parameters of this model are $y_0$ , which sets the initial value of $y$, at which the quantum evolution sets in and the parameter $c$ in (\ref{c1}), which describes the onset of quantum regime, both fitted to RHIC dAu data reported by BRAHMS collaboration \cite{brahms}.
 The infrared cutoff  $\mu=1$ GeV and the momentum scale range in $0 \sim 1$ GeV are fixed by lower energy data. 
 
Now Eqs. (\ref{34} - \ref{333}) that define the nuclear configurational entropy are used to compute the cross section saturated component (\ref{saat}).  After awkward 
computations, the results are presented in the following figure:
\begin{figure}[!htb]
       \centering
                \includegraphics[width=.98\linewidth]{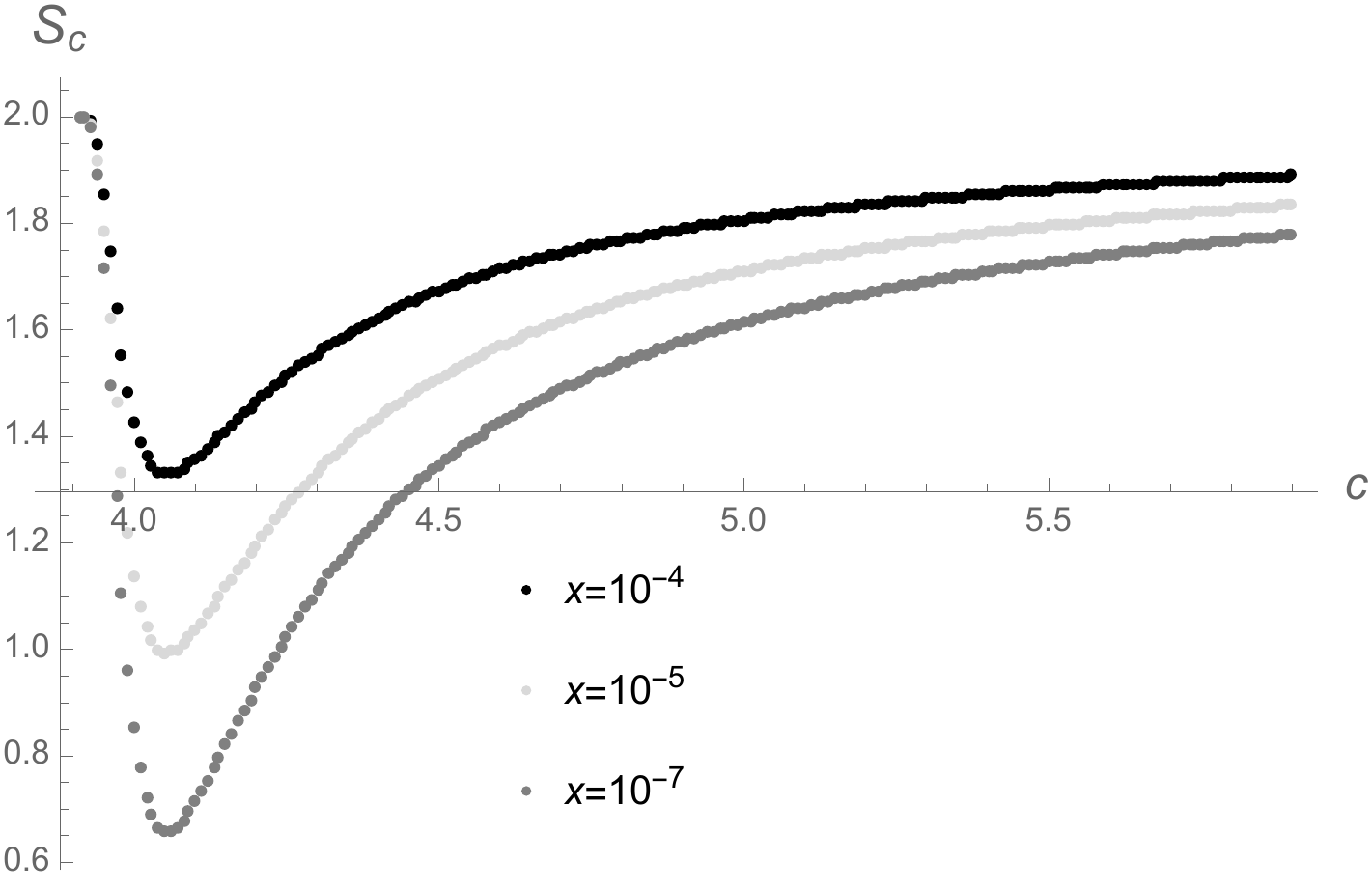}
                \caption{Configurational entropy as a function of the onset of the gluon anomalous dimension, in the Color-Glass Condensate regime, for distinct values of the Bjorken variable $x$. Each curve was generated for intervals of $c=0.01$. The black curve 
is depicted for $x=10^{-4}$, the light grey curve is plot for $x=10^{-5}$ and 
the grey curve represents $x=10^{-7}$. }
                \label{curv-ricci}
\end{figure}
The informational entropy computed by Eq. (\ref{333}) is shown in Fig. 1, for different values of the Bjorken variable $x$. Small $x$ values are employed, 
according to Refs. \cite{kkt,Carvalho}. 

The minima in the curves that depict the informational entropy  at Fig. 1 
occur at $c=4.038$, for $x=10^{-7}$; at $c=4.041$, for $x=10^{-5}$; and  at $c=4.050$, for $x=10^{-4}$. The fitted parameter $c$ establishes the onset of quantum regime, and was assumed $c=4$, in Refs. \cite{kkt,Carvalho}, to derive the results therein presented.

\subsection{Conclusions}
Here we showed that such is a natural choice provided 
by the analysis of the minima of the informational entropy, associated to the  
the onset of the gluon anomalous dimension, in the Color-Glass Condensate regime. For the Bjorken variable $x=10^{-7}$, the error is $\sim0.94\%$, whereas for $x=10^{-4}$ the error is $\sim 1.23\%$. This is the supremum 
among the error for any curve for fixed Bjorken variable. Indeed, 
 dense systems of gluons of small $x$ represent the Color-Glass Condensate based on QCD \cite{cgc}, and the Bjorken variable  $x_0=0.3 \times 10^{-4}$ was  fixed by initial conditions. Hence, all value for the Bjorken variable must be adopted as being smaller then this initial condition. 
A direction to be further studied encompasses quantum mechanics fluctuations, regarding Eqs. (\ref{saat}, \ref{satt}). The wave function used 
in those equations can be explored in the context of topological defects proposed, e. g., in Refs. \cite{Bernardini:2016rgb,Bernardini:2016uhj,German:2012rv}, in the configurational entropy setup.
Moreover, a modified  KKT model \cite{kkt} uses a modification of the
KKT model assuming that the saturation momentum scale is given by $y_0 =4.6$ and the onset of quantum regime can be taken a different value \cite{Carvalho,Carvalho:2015eia,Capossoli:2015sfa}. 
  \acknowledgements
GK thanks to FAPESP (grant No. 2016/18902-9), for partial financial support.

\end{document}